\documentclass{aa}  
\usepackage{natbib}
\usepackage{graphicx}
\usepackage{color}
\usepackage{txfonts}
\usepackage{mathtools}
\begin{document} 

  \title{Photospheric plasma and magnetic field dynamics during the formation of solar AR 11190}
\titlerunning{Plasma and magnetic field dynamics during the formation of AR 11190}


   \author{J. I. Campos Rozo
          \inst{1,2}
          \and 
           D. Utz\inst{1,3}
          \and
          S. Vargas Domínguez\inst{2}
          \and
          A. Veronig
          \inst{1}
          \and
          T. Van Doorsselaere
          \inst{4}
          }

   \institute{IGAM, Institute of Physics, University of Graz, Universitätsplatz 5, 8010 Graz, Austria\\
              \email{jose.campos-rozo@uni-graz.at}
         \and
         Universidad Nacional de Colombia, Observatorio Astronómico Nacional, Ed. 413 Bogotá, Colombia\\
             \email{svargasd@unal.edu.co}
             \and
         Instituto de Astrofísica de Andalucía IAA-CSIC, Es-18008 Granada, Spain\\
           \email{dominik.utz@uni-graz.at}
             \and
             Centre for Mathematical Plasma Astrophysics, Department of Mathematics, KU Leuven,
3001 Leuven, Belgium\\
             \email{tom.vandoorsselaere@kuleuven.be}
             }

   \date{Received XXX ; accepted XXX}

 
  \abstract
   {The Sun features on its surface typical flow patterns called the granulation, mesogranulation, and supergranulation. These patterns arise due to convective flows transporting energy from the interior of the Sun to its surface. The other well known elements structuring the solar photosphere are magnetic fields arranged from single, isolated, small-scale flux tubes to large and extended regions visible as sunspots and active regions.}
   {In this paper we will shed light on the interaction between the convective flows in large-scale cells as well as the large-scale magnetic fields in active regions, and investigate in detail the statistical distribution of flow velocities during the evolution and formation of National Oceanic and Atmospheric Administration (NOAA) active region 11190.}
   {To do so, we employed local correlation tracking methods on data obtained by the Solar Dynamics Observatory (SDO) spacecraft in the continuum as well as on processed line-of-sight (LOS)  magnetograms.}
   {We find that the flow fields in an active region can be modelled by a two-component distribution. One component is very stable, follows a Rayleigh distribution, and can be assigned to the background flows, whilst the other component is variable in strength and velocity range and can be attributed to the flux emergence visible both in the continuum maps as well as magnetograms. Generally, the plasma flows, as seen by the distribution of the magnitude of the velocity, follow a Rayleigh distribution even through the time of formation of active regions. However, at certain moments of large-scale fast flux emergence, a second component featuring higher velocities is formed in the velocity magnitudes distribution.}
   {The plasma flows are generally highly correlated to the motion of magnetic elements and vice versa except during the times of fast magnetic flux emergence as observed by rising magnetic elements. At these times, the magnetic fields are found to move faster than the corresponding plasma.}

   \keywords{Photosphere --
               granulation --
               evolution --
                magnetic fields --
                sunspots
               }
 
   \maketitle
%

\section{Introduction}
High-resolution observations have shown that the solar photosphere is a non-uniform layer formed by different structures that are constantly evolving at multiple spatial and temporal scales. Some of these features form patterns, so-called convective cells, granules, ``or in more colloquial terms, bubbles'' \citep{Hansen2004}. In the centre, these convective cells feature the emergence of hot plasma with vertical velocities of about $0.4$~km~s$^{-1}$, while towards their boundaries, the plasma motions become horizontal, moving to the intergranular lanes with velocities of $0.25$~km~s$^{-1}$ \citep{Title86}. Diverse studies have shown how granular cells can organize themselves in three main size scales: granulation \citep[$\sim 1000$ km;][]{Bushby2014}, mesogranulation \citep[$5 - 10$ Mm;][]{November1981}, and supergranulation \citep[> 20 Mm;][]{Rieutord2010}. 

\begin{figure*}

\centering
\includegraphics[width=18.5cm]{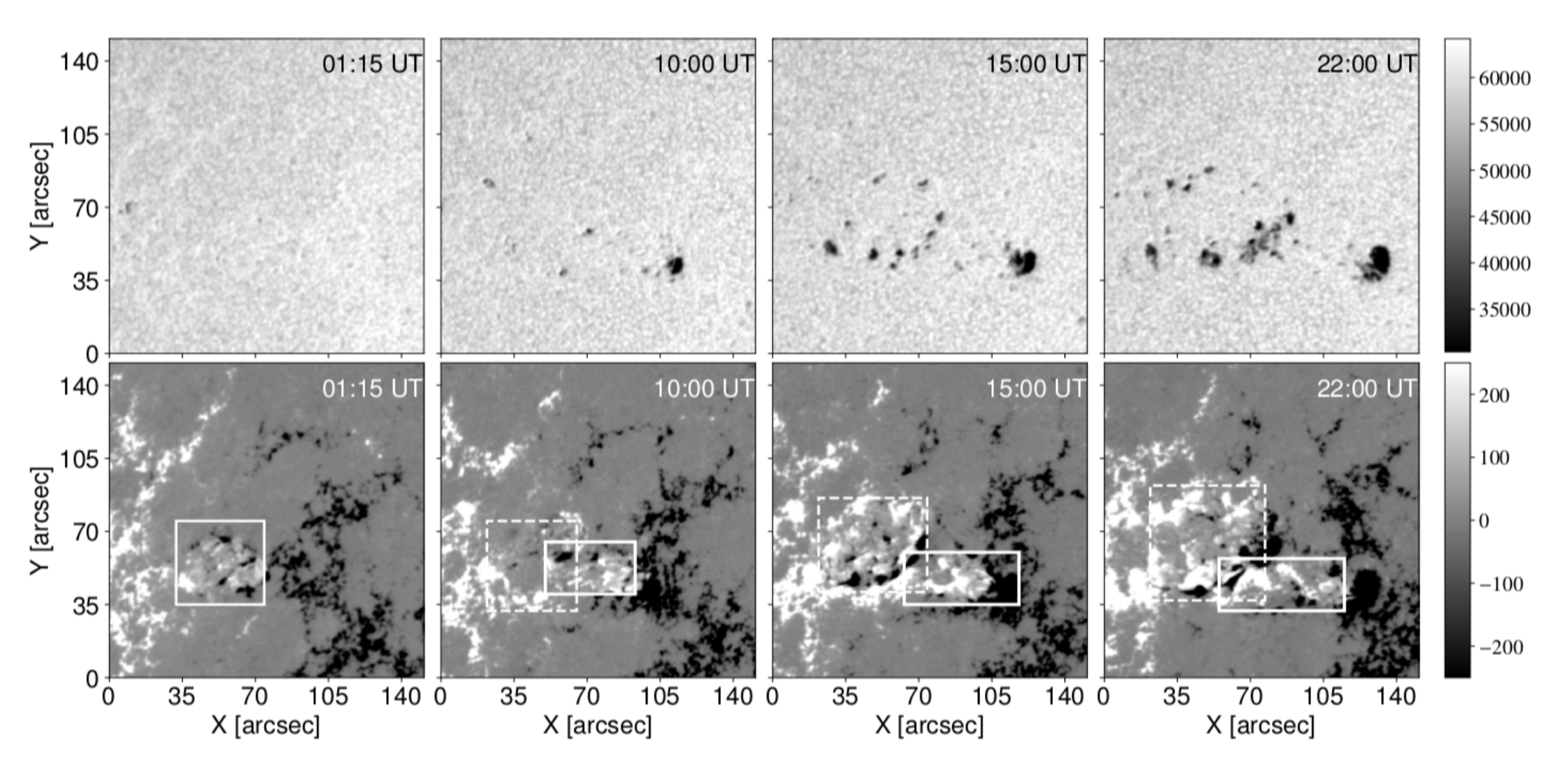}
\caption[Caption for LOF]{Selected active region (AR) as observed on April 11, 2011 exhibits a complex configuration with sunspots harbouring partial penumbrae and pores. The first row shows the time evolution of the continuum images for AR 11190 that displays a rapid evolution. The second row plots series of LOS magnetograms for AR 11190 with positive and negative (white  and black) magnetic polarities used to track the evolution of emergent positive magnetic field. The solid box encloses the region of an initial positive magnetic emergence, while the dashed box outlines a second emergent cell evolving faster and thus pushing and suppressing the first emergence. The colour bar in the first row displays the intensity values clipped between 45\% and 95\% of the image featuring the largest intensity, whereas the colour bar in the second row shows the LOS magnetic field maps clipped in the range between $-250$ and $+250$ Gauss. A clearer understanding can be obtained by watching the full evolution of the active region in the  movie provided, which shows the magnetic field on a false colour table from $-2500$ G to $+2500$ G, and the continuum maps normalized over the average of all the maximum (additional online supplement to the paper).\footnotemark}
\label{fig1}
\end{figure*}

The evolution of the photospheric granulation pattern is determined by the expansion and the subsequent dissolution of granular cells \citep{Rezaei2012, Palacios2012}. When granules start to show the appearance of bright rings, a dark centre, and fast outflows, they are called exploding granules \citep{Title86,Title1989}. This happens during the final expansion at the last stage of their life. Such explosive granules are considered important in the context of the whole solar granulation evolution \citep{Rosch1961,Musman1972,Namba1977}. Understanding the behaviour of the different granulation scales and granular evolution is important in order to obtain more realistic quiet Sun models for the formation, as well as the evolution and decay of active regions \citep{Roudier2003}. The physical processes of the emergence of granular-scale magnetic fields are likely to be the same in the quiet Sun and active regions \citep{Vargas2012}.

\begin{figure*}[]
\centering
\includegraphics[width=18cm]{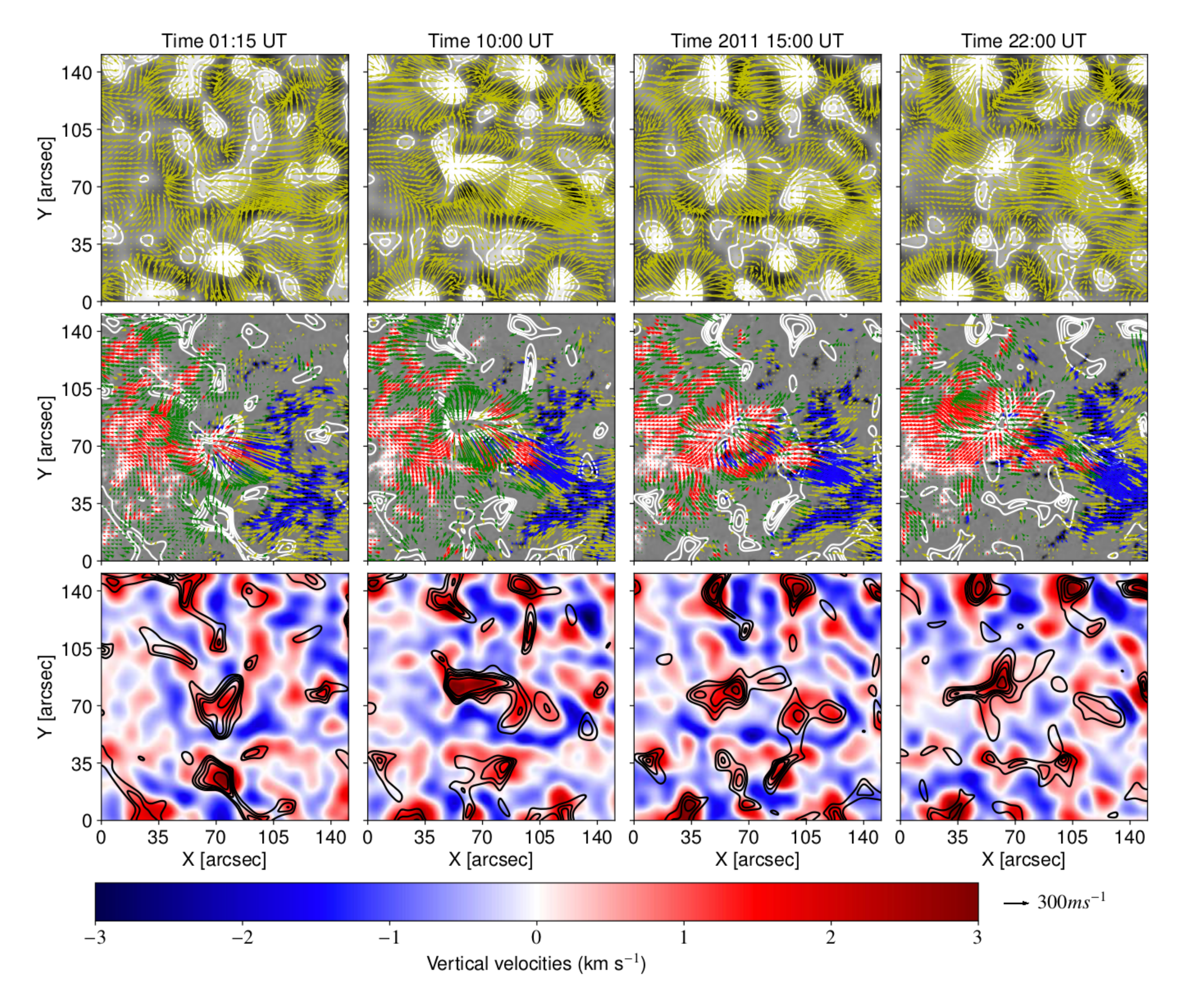}
\caption{Temporal evolution of the velocity fields within the ROI of AR 11190 at four different times computed by local correlation tracking (LCT) analysis. Horizontal and vertical velocities are inferred from continuum maps as well as from LOS magnetic field maps. The top panel shows the horizontal proper motions (calculated by LCT technique applied to the continuum maps) with the background image being the vertical velocity map, and the contour lines representing positive vertical velocities with contour values of [0.5, 1., 1.5, 2., 3] km s$^{-1}$. The second row displays the horizontal velocities of magnetic elements (calculated from LCT analysis over the LOS magnetograms) where the background image represents the LOS magnetic field strength map and the contour lines are ranged as before. The red arrows show the motions of magnetic elements with magnetic strengths greater than 50 Gauss, and the blue arrows display the average movements of negative magnetic elements with values lower than -50 Gauss. The green arrows display horizontal behaviour for weak positive magnetic elements, whereas the yellow arrows display the horizontal proper motions associated with weak negative magnetic field elements. The third row shows a comparison between the evolution of vertical velocities obtained from the continuum data set and the evolution of positive vertical velocities obtained from the LOS magnetic field data. The black arrow in the bottom right corner represents the length of a velocity  vector featuring a magnitude of $300$ m s$^{-1}$.}
\label{fig3}
\end{figure*}

Large-scale granules (mesogranules and supergranules) have been associated to fast vertical upflows \citep{Roudier2003, Guglielmino2010, Palacios2012, Verma2016}. Moreover, exploding granules occur with certain preferences in mesogranular regions \citep[e.g.][]{Massaguer1980,Title1989} and during the emergence of magnetic field and its subsequent accumulation within mesogranular borders \citep{Simon1988,Dominguez2003,Ishikawa2011}.\\
In this paper, we focus on the behaviour of such large-scale granular cells within the region of interest (ROI) and at the time when National Oceanic and Atmospheric Administration (NOAA)  active region 11190 is formed. Such detailed studies were not possible, due to the lack of highly resolved, long, and sufficiently stable  time series for the detailed investigation of the formation of an active region over days on spatial scales down to the granulation. With highly sophisticated space instruments like the SDO spacecraft \citep[Solar Dynamics Observatory, see][]{Pesnell2012}, used in this study, we can follow the long temporal evolution necessary for the formation and dissolution of whole active regions. 
\footnotetext{A movie showing the full evolution can be found at \url{https://github.com/Hypnus1803/PlotsSunspots/blob/master/VideoPaper/CubeMag_20110411_hmicmap.mp4}.}

For the detailed flow investigations performed in this study, we need to identify the proper surface motions. Several authors have studied these kinds of proper motions using different approaches \citep[e.g.][]{Hurlburt1995,Simon1995,Welsch2004,Schuck2006}. Among the most widely applied methods is the so-called local correlation tracking (LCT) technique. LCT was employed for the calculation of proper motions of the solar granulation for the first time by \citet{November1986}, and subsequently by \citet{November1988}. The algorithm is based on finding the best cross-correlation between two consecutive images. It is well known that LCT can produce some errors due to the method itself when it measures the intensity changes as these changes can be related to plasma motions but also may reflect phase velocities  \citep[e.g.][]{Roudier1999,Potts2003}. In addition, LCT may produce errors through phenomena such as the shrinking-sun effect \citep{Lisle2004} caused by large stationary flows. However, one can correct for these errors and thus such artefacts can successfully be removed by subtracting the time average of the velocities from the flow maps. Some authors \citep[e.g.][]{Yi1992,Molowny1994,Verma2013,Louis2015,Asensio2017} have shown that in the worst cases the underestimations of velocities may amount to 20-30\%.

\section{Data and pre-processing}

\begin{figure*}

\sidecaption
\includegraphics[width=12.5cm]{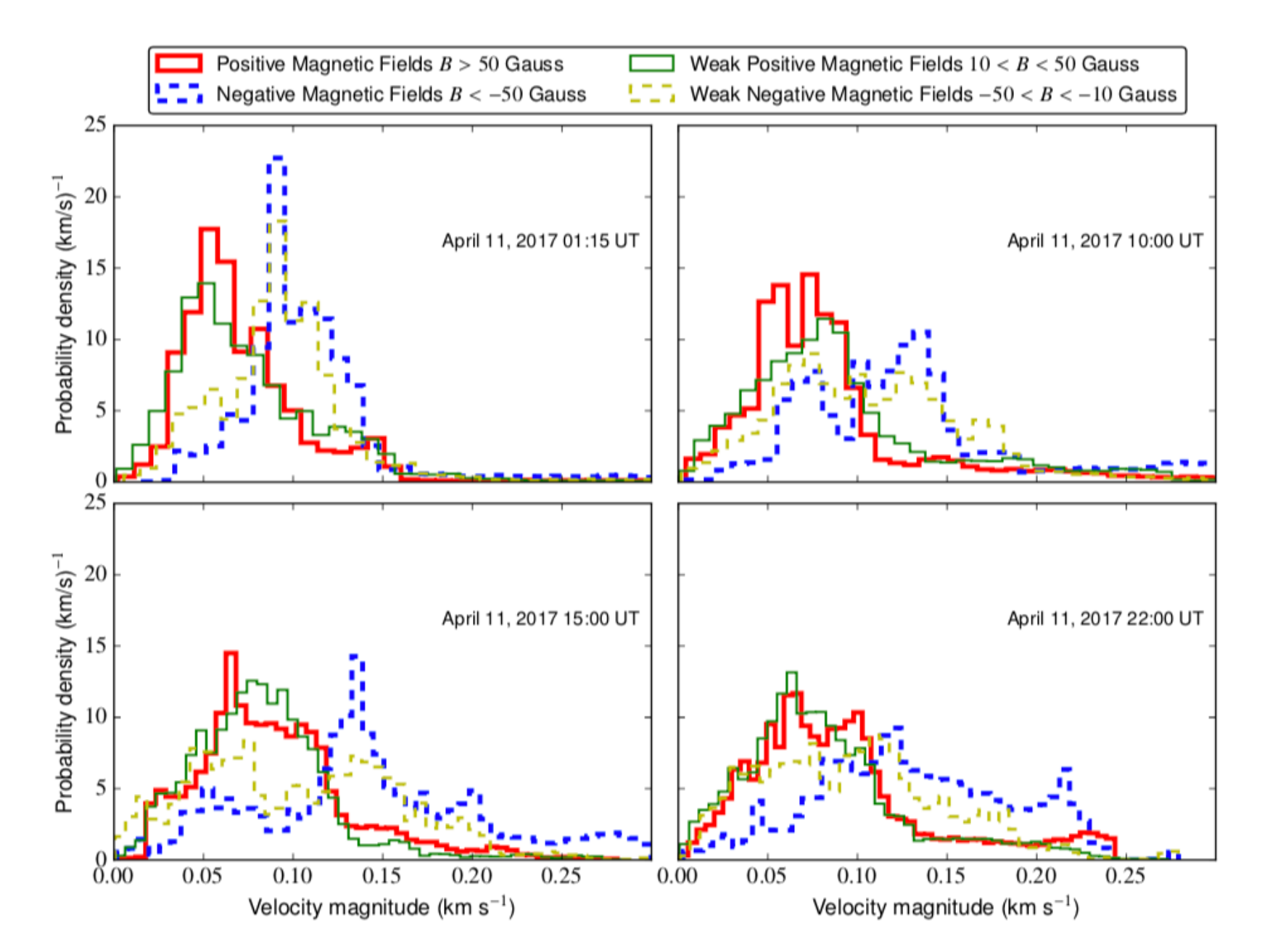}
\caption{Distribution of magnitudes of velocities ($v=\sqrt{v_x^2+v_y^2}$; speeds) for weak and strong magnetic fields for the times shown in Fig. \ref{fig3}. Red and blue colours (thick lines) describe the velocity distributions for positive and negative magnetic fields greater than $50$ Gauss, whereas green and yellow (thin lines) distributions represent the motions of the weaker magnetic elements. The ranges for both the weak positive and negative field strengths are [$10<B<50$ \& $-50<B<-10$] Gauss. Here, $B$ is the magnetic field strength as obtained from the magnetograms. The solid lines represent the positive magnetic polarity, whereas the dashed lines represent the negative magnetic fields.}
\label{fig4}
\end{figure*}

In this study we use time series of imaging data showing the formation and evolution of active region (AR) 11190. The data were acquired by the Helioseismic and Magnetic Imager (HMI) instrument \citep{Hoeksema2014} on board the SDO spacecraft comprising continuum maps and line-of-sight (LOS) magnetograms.

\begin{figure*}

\sidecaption
\includegraphics[width=12.4cm]{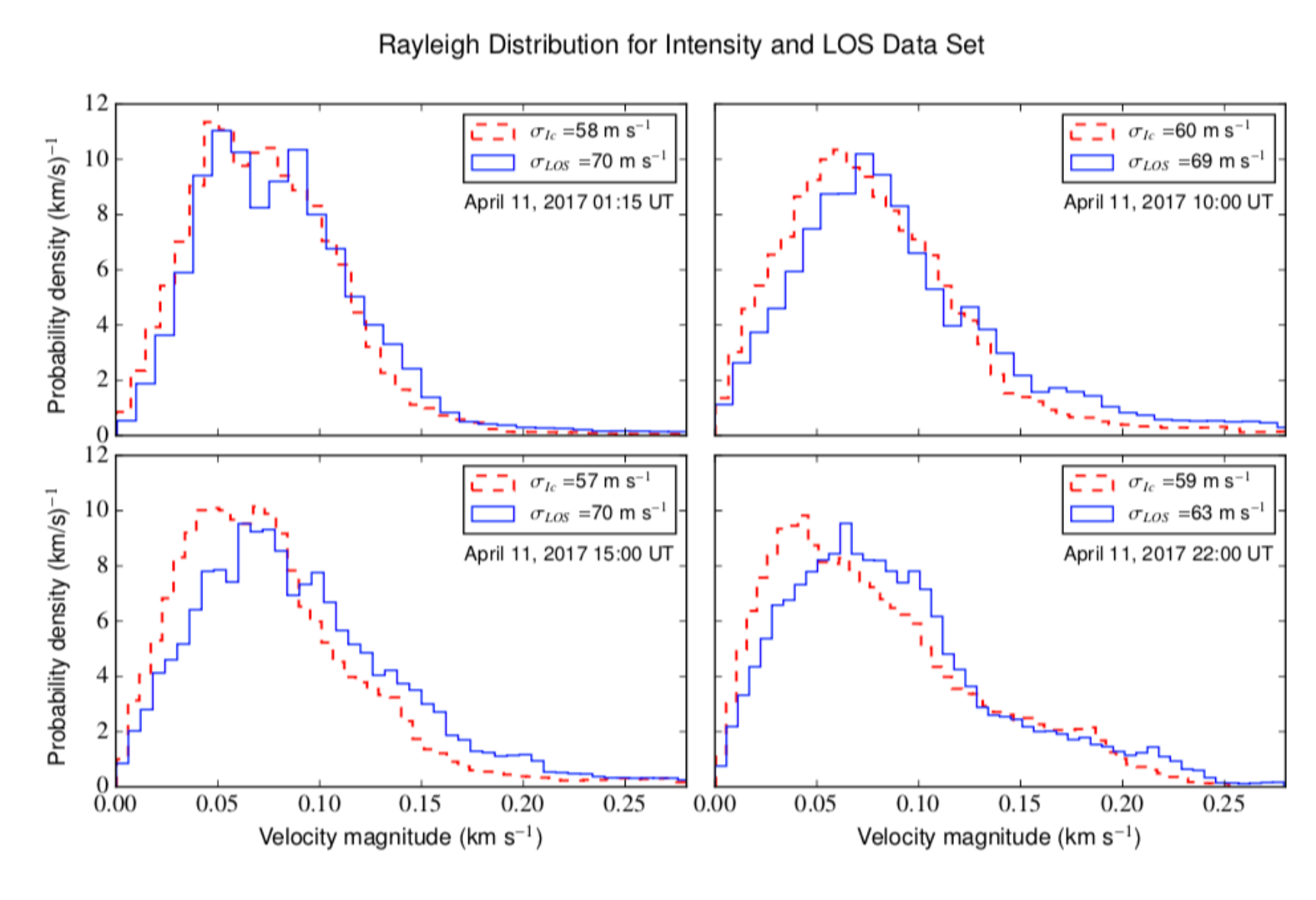}
\caption{Distribution of speeds. The red dashed line shows the distribution of $v$ for the proper motions, whereas the blue and solid line shows the distribution of magnetic elements motion. The scale factor $\sigma$ is associated with the mean velocity in this distribution.}
\label{fig5}
\end{figure*}

The studied AR 11190 is shown in Fig. \ref{fig1}, and comprises the formation stage during the time interval for the whole 24 hours on April 11, 2011 in continuum as well as in LOS magnetic field. The maps are acquired with a cadence of 45 seconds and a pixel resolution of $\sim 0.504$ arcsec (roughly 350 km) for both observable quantities (continuum, LOS magnetic field). As these data come already prepared at level 1.5, no primary data reduction such as flat fielding and dark current correction is necessary.\\

Figure \ref{fig1} shows four different instances prior to the formation of AR 11190 in continuum and LOS magnetic field images. The video linked to the web url mentioned above shows clearly the emergence of the first magnetic bubble \citep{Ortiz2014,Jaime2015}, and then a second even faster and more powerful magnetic emergence starts to occur, lifting more positive magnetic flux to the surface and pushing the previously emerged flux to the right. 
The faster emergence is seen even more clearly in the evolution of magnetograms compared to the observations in the continuum data. In this LOS magnetogram, a solid box (see second row in Fig. \ref{fig1}) encloses an initial positive magnetic field emergence, which starts during the first hours of the day, whereas a dashed box encloses the second magnetic emergence. The LOS maps displayed in the figure were clipped in the range [$-250,250$] Gauss.\\

Moreover, Fig. \ref{fig1} (LOS maps) shows pre-existent positive and negative magnetic field regions. When the first magnetic bubble appears within the field of view (FOV), the positive and negative magnetic elements further away from the site of emergence (constituting in some way the background magnetic environment) do not change noticeably. At the same instant that the second magnetic bubble starts to emerge, it pushes the first emergence away from the location of newly emergent flux. Thus, soon after, all the previously emerged magnetic elements are pushed in the same right direction towards the negative magnetic elements. These magnetic elements in turn start to accumulate at certain locations, increase there in magnetic flux, and evolve for the first time into small magnetic pores that later on  become the fully evolved AR 11190.

\begin{figure*}

\centering
\includegraphics[scale=0.7]{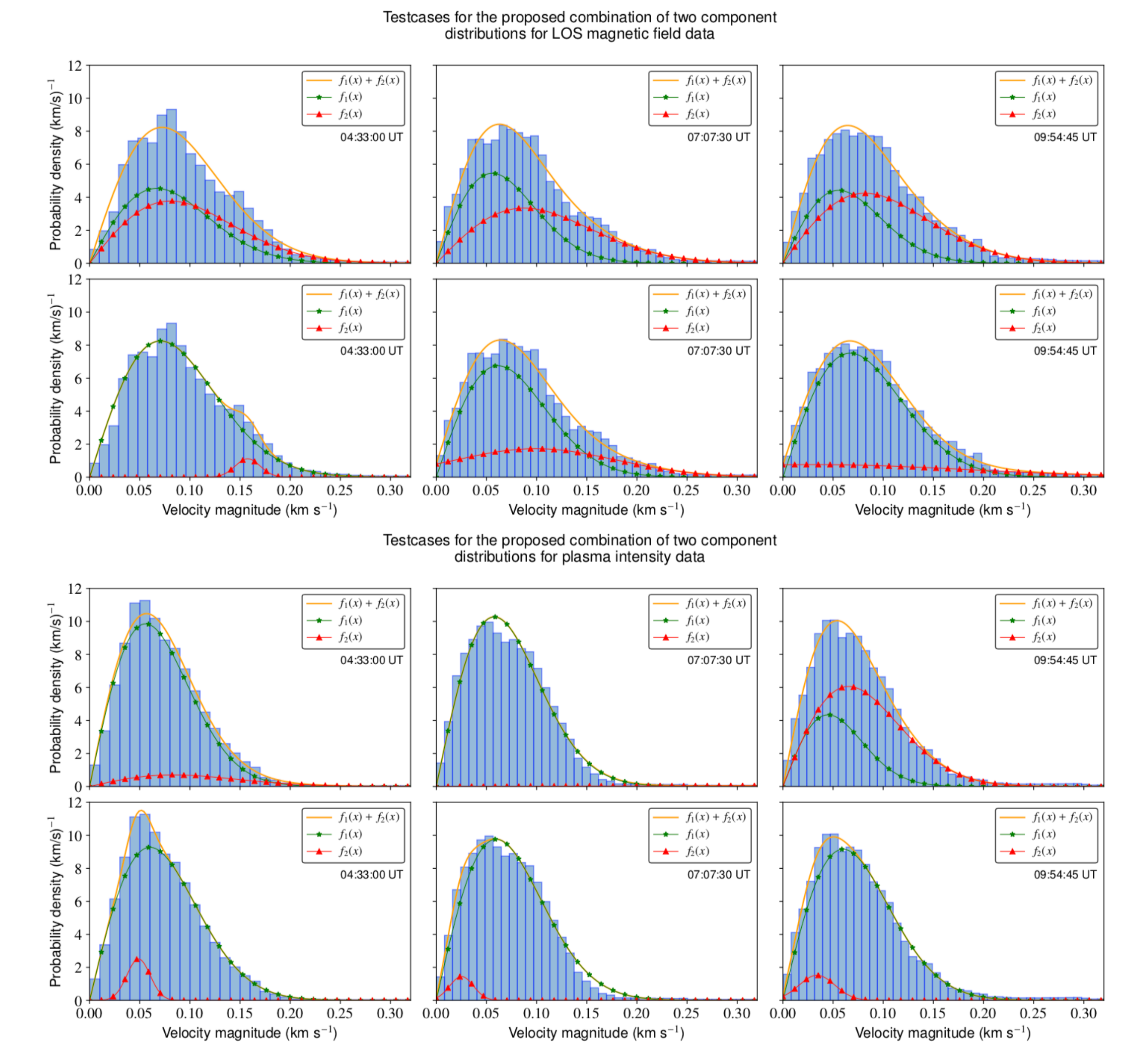}
\caption{Plots show three test cases for the proposed combination of two-component distributions for the {continuum} flow maps (lower set of panels) as well as for the magnetic elements motions  (upper set of panels). The first row in both six panel sets depicts the case of combining two Rayleigh distributions for the total fit. They  display the two independent Rayleigh components as well as the sum of the two components to form the whole measured velocity distribution for three different times. The second row for each of them displays in the same way the case of the combination of a Rayleigh background component with a variable Gaussian component. Times are in universal time (UT).}
\label{fig6}
\end{figure*}

The software used for reducing HMI data (e.g. derotation, coaligment, and subsonic filtering procedures) was encoded in the Python language making use of the solar physics library named Sunpy \citep{Sunpy2015}. A graphical user interface (GUI) has been developed to facilitate the detection and application of the LCT method \citep[see ][]{Campos2014}\footnote{The code can be found at \url{https://github.com/Hypnus1803/FlowMapsGUI}.}. The ROI was chosen manually in such a way as to centre on the location where the emergence of fast and highly notable large-scale granules is happening. The size of the analyzed FOV is $150\arcsec\times 150\arcsec$. All images were aligned and a subsonic filtering with a phase-velocity threshold of $4$ km s$^{-1}$ was applied to subtract the solar 5-minutes oscillation \citep{November1981,Title1989}. Moreover, due to the Sun being a hot plasma sphere and the sunspot locations spreading along different regions on the solar disc, flow map velocity components were properly deprojected \citep[see ][ and references therein]{Vargas2009}.

\section{Results}
We focused on the formation and emergence of AR 11190, and investigated in detail the evolution and behaviour of the plasma and  magnetic field dynamics from horizontal and vertical velocities for different time ranges. The LCT technique applied is based on Eq. \ref{eq1} proposed by \citet{November1988},
\begin{equation}
C_{t,t+\tau}(\pmb{\delta},{\bf x}) = \int  J_{t}\bigg{(}\pmb{\zeta}-\frac{\pmb{\delta}}{2}\bigg{)} J_{t+\tau}\bigg{(}\pmb{\zeta}+\frac{\pmb{\delta}}{2}\bigg{)}W({\bf x}-\pmb{\zeta})\partial\pmb{\zeta},
\label{eq1}
\end{equation}
where $C_{t,t+\tau}(\pmb{\delta},{\bf x})$ is a four-dimensional function depending on two consecutive images, the displacements between these images, and the localization of the apodization window $W({\bf x})$; and $J_{t},\, J_{t+\tau}$ are the intensity of the images at two consecutive time steps $t$ and $t+\tau$. It is worth mentioning that the velocities estimated by LCT are not exclusively plasma motions but strictly speaking horizontal proper motions as the algorithm does not use plasma physical properties.
The LCT algorithm applied in this work \citep[see ][]{Yi1992,Molowny1994} was adapted in Python to calculate the velocity fields using an apodization window adjusted for the comparable size of the features to be tracked. Authors such as \citet{Palacios2012} have shown that the emergence of new magnetic flux as well as posterior AR formation are associated with explosive mesogranules. 

For that reason we have chosen a full width at half maximum (FWHM) parameter of 12.5 arcsec ($\sim 9$ Mm) corresponding to typical average sizes of ensembles of granules forming the meso-granular pattern, and a temporal averaging period of 2 hours \citep[average lifetimes for large-scale granulation patterns; see ][]{Hill1984,Rast2003}. These 2 hours correspond to 160 frames in the data set.\footnote{We wish to remark that given UT times on the images always correspond to the first image of such 160 images containing subsets.} Vertical velocities are computed  by the divergence from the horizontal velocities $v_x$ and $v_y$ obtained by the LCT algorithm via the idea of flux conservation \citep[see ][]{November1988,Marquez2006, Vargas2009} leading to the expression
\begin{equation}
v_z(v_x, v_y) = h_m\nabla\cdot\vec{v}_h(v_x, v_y), 
\end{equation}
where $h_m$ is a constant of proportionality representing the mass-flux scale height with a value of $150\pm 12$ km \citep[see ][]{November1987, November1989}. The flow maps are then plotted over these vertical velocities obtained from continuum maps as well as from magnetograms.

\subsection{Horizontal and vertical flow maps}
The studied AR shows exploding mesogranules in locations where the formation of the active region, as seen by a complex sunspot group, is initiated. There is a strong connection between the appearance of these emergent large-scale granules and rapid vertical upflows emerging from the same region.
Even when AR 11190 does not show strong emergences in the continuum maps, a strong emergence of positive magnetic field elements can be clearly observed in the magnetograms. Horizontal and vertical flow maps of proper motions as well as of magnetic field elements were calculated with the LCT algorithm to link the photospheric plasma dynamics with the magnetic field evolution. \\

Figure \ref{fig3} shows the evolution of the magnetic flux and plasma emergence during the formation of AR 11190 as well as the behaviour of the plasma and the movement of the magnetic elements as observed in the LOS magnetograms at four different time. The horizontal and vertical velocities are plotted in each panel showing the behaviour and giving information about the proper motions of the plasma and the magnetic elements during the appearance of AR 11190.\\
The first panel displays the evolution and behaviour of the continuum maps. The horizontal velocities are represented by the arrows overplotted in the ROI, whereas the vertical velocities are represented by the background image. These velocities reveal several divergences at the following positions: ($x\arcsec$, $y\arcsec$) = ($70$, $30$), ($30$, $5$), ($70$, $70$), or ($50$, $130$).\\
We focus now on the emergence centred on the position $70\arcsec\times 70\arcsec$ as it displays a comparably more rapid emergence of strong, as well as weak, positive magnetic field elements, as evidenced in the second row at the same location, which can be found at the other emergence sites. Although the other emergences indicate a certain correspondence of vertical motions between plasma and the LOS magnetic elements, they do not display horizontal motions of positive magnetic elements greater than 10 Gauss (lower limit used in the present work) emerging in those regions.\\
Before the appearance of the second emerging bubble, the motions of the magnetic elements follow the paths imposed by the plasma horizontal motions as well as the up- and down-flows. When the second magnetic emergence starts to appear, the proper motions seem to follow the new paths imposed by the new strong positive magnetic field elements, displaying a preferred motion in the positive x-direction (see also additional online movies). In the first row of Fig. \ref{fig3}, the contour lines show positive vertical velocities enclosing magnitudes of [0.5, 1., 1.5, 2., 3] km s$^{-1}$ calculated from the continuum data set. 

The second row in Fig. \ref{fig3} features plotted contours of the vertical velocities calculated from the LOS magnetic field data using the same contour values mentioned before. This panel shows also the horizontal motions of positive and negative magnetic elements in the LOS maps. It is possible to identify the moment when the second magnetic emergence starts to appear (row=2, column=2). This emergence shows fast motions but only associated with weak positive magnetic field elements that turn later into strong magnetic field elements.\\

In order to compare the dynamics of plasma and magnetic elements, Fig. \ref{fig3}, third row, shows in the background the vertical velocities from the continuum data set as well as overplotted contour lines representing the positive vertical velocities calculated from the LOS magnetic data cube. Although all vertical velocities calculated from magnetic LOS elements are linked to plasma vertical velocities, the best observational correlation is registered for the emergence located in $70\arcsec\times 70\arcsec$. Both flow patterns seem to evolve at the same rate and look alike. 

However, to have an even more robust and quantitative overview of the ongoing and evolving flows, we will now have a detailed look at the distribution of the magnitudes ($\sqrt{v_x^2+v_y^2}$; hereafter called speed) separated for the previously mentioned strong and weak fields, as well as for the two polarities.\\
The resulting distribution of speeds for the same four time instances (at the beginning of the first emergence, during the second emergence, after the second emergence, and to the end of the evolution) can be seen in Fig. \ref{fig4}. At the beginning of the first emergence (left upper panel) one can see a separation of the positive and negative polarities.\\ 
While the distributions for the positive magnetic field elements look more or less Rayleigh distributed (indicating a two dimensional freely, i.e. randomly, outflowing region), the same distribution for the negative magnetic elements features the appearance of a normal distribution, but offset from zero by a certain constant velocity, indicating a movement leading to a separation for the two polarities, where the negative magnetic elements tend to move towards the right side of the FOV. During and just after the second emergence (panels 2 and 3 in Fig. \ref{fig4}) the distributions seem to be truncated and merging at horizontal velocities of around 0.12 km s$^{-1}$. This behaviour can be explained by the idea that the created positive magnetic elements catch up with the negative elements towards the right side. After catching up, these negative elements then hinder the positive ones in moving faster. Thus the positive distribution gets truncated at higher velocities while the distribution for negative elements becomes truncated for low velocities as the positive elements push into the slowest negative ones thus either accelerating them to the same speed or annihilating  them when they catch up. The last panel in Fig. \ref{fig4} (lower right one) shows the evolved FOV where both kinds of magnetic elements seem to approach very similar distributions and thus move and evolve together again. \\
Figure \ref{fig5} displays the speeds of the horizontal proper motions (red dashed line) and the horizontal movements of magnetic elements (blue and solid line; now regardless of their polarity and strength). The upper panels in Fig. \ref{fig5} show a correspondence between the plasma and magnetic field distributions, which means that both are moving following the same behaviour. As they are evolving (bottom panels), the velocity distribution of the magnetic elements shows an increase in its mean value, whereas the mean velocity obtained from the proper motions appears to decrease in value most likely being suppressed by the stronger magnetic fields. Due to the physical processes creating the flows, the best description of the distribution of speeds is generally given by a Rayleigh distribution (Eq. \ref{eq2})\footnote{Mathematically, a Rayleigh distribution for the magnitude of a two-dimensional vector is formed when both vector components follow 0-centred normal distributions with equal $\sigma$ (standard deviation), which is common for random walk processes (convective flows)}, 

\begin{equation}
f(v,\sigma) = \frac{v}{\sigma^2}\exp{\bigg{(}\frac{-v^2}{2\sigma^2}\bigg{)}},\, v>0,
\label{eq2}
\end{equation}

where the scalar factor $\sigma$ is associated with the mean velocity of such a distribution \citep{Hoffman1975}. The mean velocity for a quiet small region, using temporal averages of 2 hours, is $v_{Int}=72\pm 8.8$ ms$^{-1}$ for the continuum maps, whereas for LOS magnetogram data the value amounts to $v_{LOS}=54\pm 10.7$ ms$^{-1}$. However, when the mean velocity is calculated over the chosen ROI, the horizontal proper motions obtained from the magnetic elements data ($v_{LOS}=65\pm 2.2$ ms$^{-1}$) appear to be slightly larger than the continuum proper motions ($v_{Int}=55\pm 2.5$ ms$^{-1}$). This difference can be explained by the second faster emergence of magnetic field. During this emergence the plasma takes some time until it starts to feel the influence of these new magnetic fields that emerge faster than the first appearance and start to push the old magnetic elements. In the photosphere most of the plasma is, due to the comparably low temperatures, in a neutral state. Thus, in the beginning of the flux emergence, only the present ions will react immediately, while some time is needed to transfer the momentum from the ions to the neutral gas. Therefore the proper motions linked to the continuum maps and their velocity distributions can lag behind the distributions of the magnetic elements. Besides, we have to have in mind that the formation height for the continuum maps can be slightly different from the formation height of the magnetograms. The existence of a strong concordance between both distributions, continuum and magnetic elements motions, is nevertheless evident even though the magnetic elements move horizontally slightly faster than the horizontal motions computed from continuum maps.

\begin{figure*}
\begin{tabular}{cc}
\includegraphics[scale=0.7]{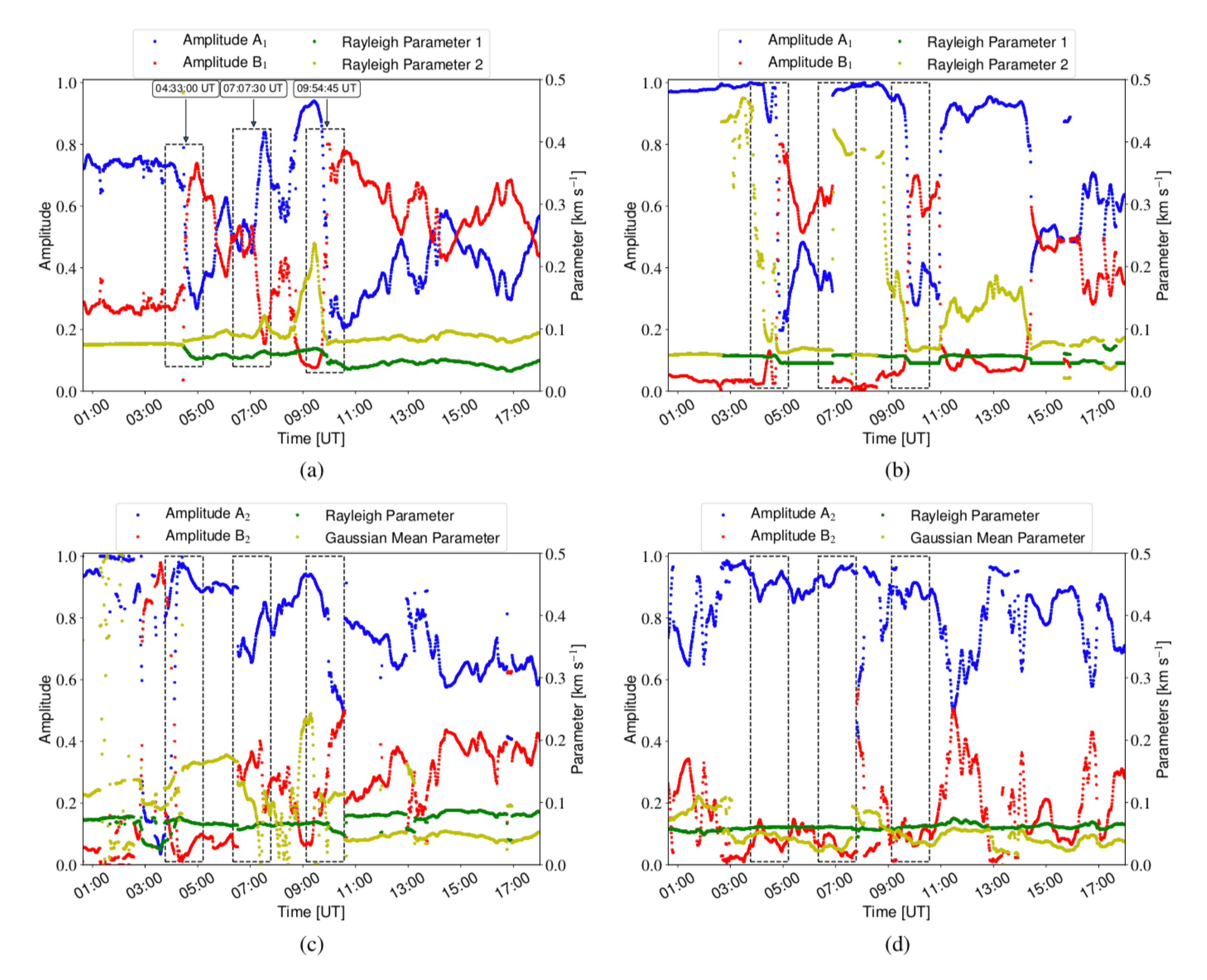} \\
 \end{tabular}
 \caption{Temporal evolution of the parameters associated with the proposed distributions in Eqs. \ref{eq3} and \ref{eq4}. The first row shows the behaviour of the parameters for the sum of two Rayleigh functions. The second row displays the time evolution for the combination of one Rayleigh function and one Gaussian function. The temporal evolution followed by the amplitudes (dimensionless -- left axis) is plotted with circle markers, whereas the temporal evolution for the functional parameters are shown with star markers (right axis in km s$^{-1}$). The blue and green markers are associated to the  first Rayleigh function component in both proposed distributions, whereas the red, and yellow markers represent the second component, which is either a second Rayleigh function or a Gaussian function. The left side of the panels are calculated from the LOS magnetograms, whereas the right side panels are obtained from the continuum maps. The dashed rectangles enclose the three time instants shown in Fig. \ref{fig6}.}
\label{fig7}
\end{figure*}

\subsection{Distribution analysis}
Figure \ref{fig5} shows, for the velocity distribution of the magnetic field elements,  distinct enhancements  variable in position as well as amplitude. Due to this behaviour, we introduce and consider a combined distribution model made up of two components. While the major part of the histogram follows a Rayleigh distribution (first component) representing undisturbed quiet background flows, the second component will be generally related to the flux emergence process creating, for example, a tail of high velocity measurements. In addition to the increased velocity tail, it is also possible that during the flux emergence a bifurcation of the velocity distribution happens due to different velocity distributions for the two magnetic field polarities, meaning that one kind of magnetic element moves with a different characteristic speed than the other. Thus the bumpy nature can be explained by the flux emergence process and/or a bifurcation of the underlying velocity distributions for the two magnetic polarities. Due to the unknown nature of the second distribution, we will employ fitting tests with a combination of either two Rayleigh distributions (see Eq. \ref{eq3}) or a combination of one Rayleigh component and one Gaussian component (see Eq. \ref{eq4}):
\begin{equation}
\begin{multlined}
f(v,\sigma_{R_1}) + f(v,\sigma_{R_2}) = \\ 
 A_1\cdot\frac{v}{\sigma_{R_1}^2}\exp{\bigg{(}\frac{-v^2}{2\sigma_{R_1}^2}\bigg{)}} + B_1\cdot\frac{v}{\sigma_{R_2}^2}\exp{\bigg{(}\frac{-v^2}{2\sigma_{R_2}^2}\bigg{)}}, 
\end{multlined}
\label{eq3}
\end{equation}

\begin{equation}
\begin{multlined}
f(v,\sigma_{R_3}) + f(v,\mu_G, \sigma_{G})  = \\ 
 A_2\cdot\frac{v}{\sigma_{R_3}^2}\exp{\bigg{(}\frac{-v^2}{2\sigma_{R_3}^2}\bigg{)}} + \frac{B_2}{\sqrt{2\pi}\sigma_G}\exp{\bigg{(}\frac{-(v-\mu_G)^2}{2\sigma_{G}^2}\bigg{)}}.
\end{multlined}
\label{eq4}
\end{equation}

\begin{figure*}[]
 \includegraphics[scale=0.7]{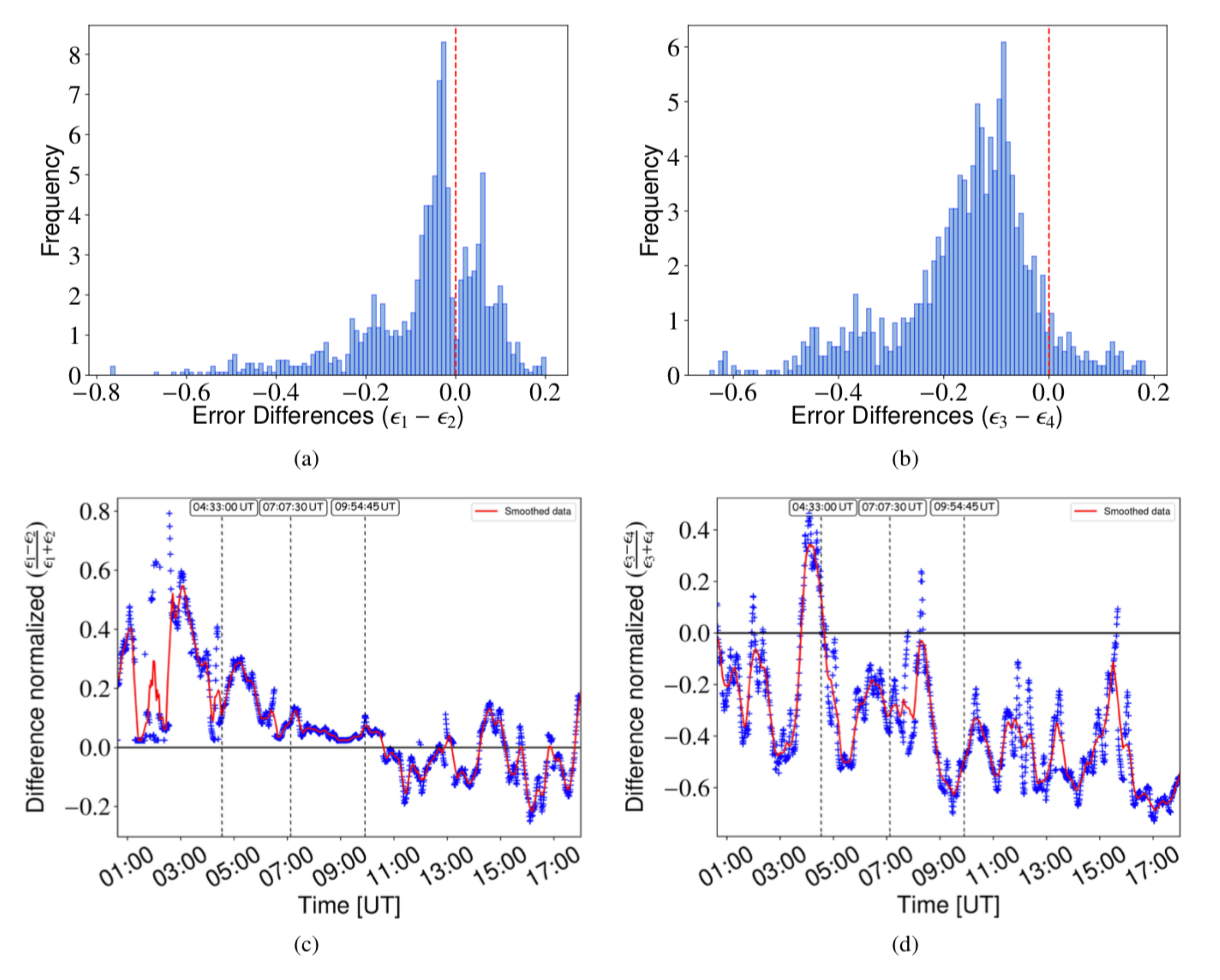}\\ 
\caption{Normalized reduced $\chi^2$ value ($\epsilon_x$) was calculated for the two fitting combinations to decide the goodness of a fit, that is, which combination of functions would fit the flow maps better. Panels a and b show the distribution for the difference between $\epsilon_1$ and $\epsilon_2$, for the distributions obtained from the LOS magnetograms, as well as for the difference between $\epsilon_3$ and $\epsilon_4$, for the distributions obtained from the continuum maps. Normalized reduced chi squared values $\epsilon_1$ and $\epsilon_3$ are related to Eq. \ref{eq3}, whereas $\epsilon_2$ and $\epsilon_4$ are related to Eq. \ref{eq4}. Panels c and d show the temporal evolution of the changes of these differences between corresponding $\epsilon$.}
\label{fig8}
\end{figure*}

In these equations $\sigma$ represents either the previously introduced scalar parameter of a Rayleigh distribution or the standard deviation in the case of the Gaussian. Constants $A_x$ and $B_x$ are the amplitudes or weighting parameters for the two components of the distribution with $x=1$ representing the double Rayleigh distribution and $x=2$ corresponding to the Rayleigh and Gaussian combined distribution. Finally, $\mu_G$ represents the mean value of the Gaussian distribution. By applying such a model, we would implicitly assume that the flux emergence leads in a part of the FOV to a secondary Rayleigh distribution or a Gaussian one most likely featuring higher velocities than the background flow distribution.

Three test cases at different time instants of such two-component modelling of the flows in the FOV of the flux emergence are shown in Fig. \ref{fig6}. The upper part of the figure shows a set of six panels created from the LCT analysis of magnetograms. The first row of these panels shows the combination of two Rayleigh components while the lower row shows the combination of a Rayleigh component and the Gaussian distribution. The lower set of panels is arranged in the same way but created from the LCT analysis of the continuum maps. These three cases were chosen visually from Fig. \ref{fig7}a at times that showed remarkable changes in the evolution of the depicted parameters. In these three test cases it becomes clear that sometimes the combination of two Rayleigh components fits better, while in other cases the combination of a Rayleigh component with a Gaussian component gives a better fit.

From these modelling efforts, we can learn that it is not straightforward and clear whether the additional component should be of Rayleigh or of Gaussian type. For instance, for the speed histograms created from the continuum maps as depicted in the first column, it is easy to observe that the combination of a Rayleigh distribution with a Gaussian distribution fits better compared to the combination of two Rayleigh distributions. This is also true for the distribution created by the LOS magnetic field data set. The middle column of distributions created from the magnetograms as well as continuum maps seem to be equally well fitted by both kinds of combined distributions, while the last column shows that the distributions would be better fitted by the two Rayleigh components combination. The second finding is that clearly the amplitude of the secondary component is variable in position as well as in amplitude.

We do not wish to introduce a model with too many free parameters and thus we will continue with these models that only comprise the mentioned two components. However, to shed more light on the goodness of these combinations, we will now investigate in more detail the temporal evolution of the parameters of such two-component models. Later, we will then also study the goodness of the fit of the combined models to ascertain which one is more likely to represent the flows in the FOV during flux emergence events.

Figure~\ref{fig7} displays an example of how the fit parameters behave during the time evolution on April 11, 2011. The left column shows the behaviour for the LOS magnetic field data set, whereas the right column gives the information about the proper motions obtained from the continuum maps. Figures~\ref{fig7}a and~\ref{fig7}b present the parameters $A_1$, $B_1$, $\sigma_{R_1}$ , and $\sigma_{R_2}$, calculated using a python least-square algorithm for the case of applying the sum of two Rayleigh distributions (Eq.~\ref{eq3}) for the LOS data set, as well as for the continuum data. Parameter $\sigma_{R_1}$ is related to the background velocity at those places of the ROI where the plasma or the magnetic elements are not affected by flux emergence. One can clearly observe in Fig.~\ref{fig7}a the existence of three strong deviations at three different times (marked dashed rectangles, and their respective times) that affect all the parameters at the same time (see also discussion above). 
It is easy to observe in Fig. \ref{fig7}a how, at the beginning of the evolution, $\sigma_{R_1}$ and $\sigma_{R_2}$ appear to have the same value, which means that the FOV is governed by a single type of background motion and is not yet affected by the first magnetic emergence. At 04:33 UT, the behaviour changes drastically showing a splitting between the $\sigma_{R_1}$ and $\sigma_{R_2}$ parameters. Moreover, the amplitude $B_1$ becomes larger than the amplitude $A_1$ giving more importance to the second component at this moment of evolution. The second time  shows a small enhancement of the Rayleigh parameter for the second component. However, in this instance the amplitude $A_1$ becomes greater than $B_1$. This change may be associated with the beginning of the second magnetic emergence. The third and the strongest change in the Rayleigh parameters related to the second distribution happens at 09:54:45 UT. At this point the second magnetic emergence becomes more active, associated with an increased number of positive magnetic field elements. Figure \ref{fig7}b, which shows the evolution of the Rayleigh combination for the continuum maps (horizontal proper motion), shows clearly several parameter jumps at the same times, although the behaviour in general of the parameters looks more chaotic. However, one can observe that the splitting between both Rayleigh parameters happens already earlier for the horizontal proper motions compared to the flows obtained from the magnetic elements. The combination between a Rayleigh and a Gaussian distribution  for the magnetic motions (Fig. \ref{fig7}c) shows in general that the Gaussian mean value is larger than the Rayleigh parameter. However, this behaviour changes after the third marked time (dashed rectangle), when the Rayleigh parameter becomes larger than the Gaussian mean. Contrary to these statements, Fig. \ref{fig7}d seems to show that in general the Rayleigh parameter governs the behaviour of the continuum horizontal proper motions except at certain times that are not obviously correlated to the changes mentioned before. In general, the amplitude $A_2$ is greater than $B_2$, implying that the contribution of this fit component to the overall speed histogram fit is marginal.

To decide which of the proposed two-component functions fits the data better, a quality test for the goodness of fitting must be done. In a first step, we compared statistically the goodness of fitting of the two models with each other. For this purpose, we obtained the normalized reduced $\chi^2$ values\footnote{Normalized means in this context that the values were normalized to the maximum chi square number obtained during the considered time evolution.} for both combinations of fitting functions, as well as for the LOS magnetic field data and continuum maps, namely the parameters $\epsilon_1$ and $\epsilon_2$, as well as $\epsilon_3$ and $\epsilon_4$ for i) the sum of the two Rayleigh functions and ii) the combination consisting of one Rayleigh and one Gaussian function, respectively. Then we subtracted the two $\chi^2$ values from each other, $\epsilon_1$ - $\epsilon_2$ and $\epsilon_3$ - $\epsilon_4$, respectively, and created a histogram plot for this difference. The result can be seen in the upper panel of Fig. \ref{fig8}.

The vertical red dashed line in Figs. \ref{fig8}a and \ref{fig8}b marks the zero line which in principle should separate the two domains of preferential fitting for the two different models. The distributions obtained from these differences between $\epsilon_1$ and $\epsilon_2$, as well as between $\epsilon_3$ and $\epsilon_4$, deduced from the magnetic field and continuum data, show two different regions. It is clearly observable in Fig. \ref{fig8}a that the zero line divides the distribution in two distinct regions. For values larger than 0, the best fitting is given straightforwardly by the sum of one Rayleigh function and a Gaussian function, whereas for values lower than 0 we would argue that clearly the best fitting can be obtained via a combination of two Rayleigh distributions for the LOS magnetic field data. Accordingly, we can see in Fig. \ref{fig8}b that the difference between $\epsilon_3$ and $\epsilon_4$ is normally distributed, and shifted to negative values indicating that in general the better fitting could be obtained by the combination of two Rayleigh components.

Figures \ref{fig8}c and \ref{fig8}d show the temporal evolution for $\bigg{(}\frac{\epsilon_1-\epsilon_2}{\epsilon_1+\epsilon_2}\bigg{)}$ and $\bigg{(}\frac{\epsilon_3-\epsilon_4}{\epsilon_3+\epsilon_4}\bigg{)}$, which can be interpreted as a ``quasi-polarization'' between both combinations and thus gives information about the times when which combination would actually fit the obtained velocity distribution better.
Figure \ref{fig8}c shows that previous to the first time instant (04:33 UT), the best fitting is given by the combination of one Rayleigh and a Gaussian distribution. After the second time instant (07:07:30 UT), even if the values are over the zero line, it is possible to argue that both combinations fit equally well. Then, after the third time instant (09:54:45 Ut), the evolution shows that the best fitting is given by the combination of two Rayleigh distributions. On the other hand, Fig. \ref{fig8}d shows that in general for the continuum horizontal proper motions, the distributions will be well fitted by the combination of two Rayleigh distributions, with the exception of some isolated short time instants that appear to coincide with the previously  outlined three special points of time we already discussed in Fig. \ref{fig7}. On this occasions it appears that the best fitting is given using the combination of one Rayleigh and one Gaussian distribution.

\section{Discussion}
In this work we have analysed time series of images displaying the formation of a solar active region in both continuum maps and LOS magnetograms. Results expose the presence of large-scale granular cells prior to the formation of the active region. 
The horizontal proper motions show strong outflows (divergences) at the same places where strong upflows can be identified. Although the calculated vertical velocities show upflows in different sectors within the FOV, these emergences do not exhibit strong and remarkable horizontal velocities. Besides, such upflows are generally only related to the appearance and motion of weak magnetic fields ($< 10$ Gauss). This is in strong contrast to the horizontal velocities detected roughly in the centre of the FOV. Here it seems as if the strong upflows are also associated to the formation of AR 11190. Generally we would like to state that the behaviour of the continuum proper motions and the magnetic field element motions  are strongly linked. This becomes also very clear in the third row in Fig. \ref{fig3}, where both the vertical velocities obtained from the horizontal velocities (background map) and the contours resulting from the positive vertical velocities obtained from the LOS magnetic field data are depicted together and display a strong correlation. This can be clearly seen by applying a Pearson correlation analysis between the vertical velocities from continuum maps and from the LOS magnetic field.
\begin{figure}[]

\centering
\includegraphics[scale=0.7]{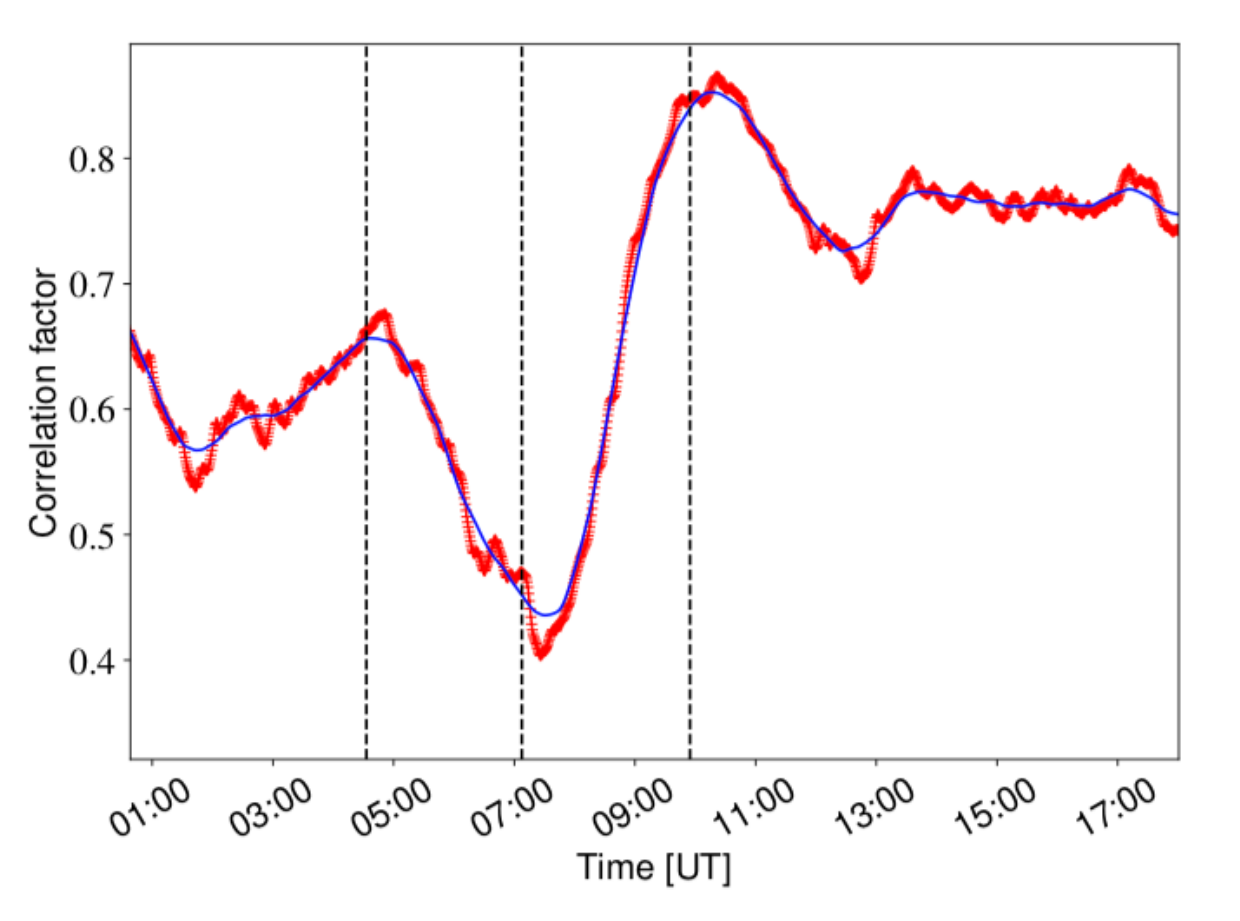}
\caption{Pearson coefficients for the vertical velocities in the region at the centre of the FOV. The size of the window was $90\arcsec\times 65\arcsec$.}
\label{fig10}
\end{figure}

Figure \ref{fig10} shows exactly such calculated correlation and its temporal evolution.
The vertical dashed lines represent the same instants as mentioned in the Fig. \ref{fig7}. All of them are close to points where the Pearson coefficient evolution changes its slope. In Fig. \ref{fig10} we can see how the correlation starts at a value of around 0.65 indicating a good correlation before dropping at 2:00 UT to a local minimum. This means that in the beginning of the time series the plasma flows and the magnetic field element motions are well coupled. However, during the first magnetic emergence, the motions become partly decoupled. This can be due to the idea that in the first moment the newly emerging magnetic field is so powerful and strong that it can weaken the coupling conditions for a moment, expand faster than the local surrounding plasma, and only slightly later start also to push away the plasma. From about 3:00 UT to 4:00 UT the coupling between the continuum proper motions and magnetic element motions are gradually restored, just to be broken again and even more strongly by the second emergence just after the first time instant mentioned above. This time the coupling gets even more weakened indicating a stronger emergence and stronger magnetic fields. Subsequently, after the second time instant, this leads in the aftermath to a stronger coupling of the continuum and magnetic field element motions most likely due to the governance of the magnetic field over the plasma owing to the strong emergence. After the third time instant, the flows become more stable and reach a high plateau of coupling with a correlation coefficient of up to 0.75. Thus we would conclude that, at least during the emergence of the second magnetic bubble, the magnetic field was strong enough to govern the plasma flows, while in other cases normal convection might advect smaller magnetic elements.

Moreover, we can see in Fig. \ref{fig10} that the maximum of correlation happens co-temporarily with the large parameter deviations as shown in Fig. \ref{fig7}a, identifiable in velocity distributions calculated for the magnetic field elements indicating again that the flow field is changing at these moments due to the emergence of new magnetic flux via the magnetic bubbles. 

As we have seen and outlined in the results section, the velocity distributions, obtained from the flow maps, are well represented by the combination of two separate components. The proposed two different combinations as given by Eqs. \ref{eq3} and \ref{eq4} feature both a Rayleigh distribution, which we would think of as fitting the undisturbed background flows, that is, the regions of the FOV not affected by the flux emergence event, plus a second component, which is variable in position and relative strength as it applies for the occupied and effected area of the flux emergence process.\\
The open question to settle is, which one of the proposed components generally fits better? The Gaussian or the Rayleigh distribution? We believe that it might not be clear as there could be a kind of phase transition between both distributions. A Rayleigh distribution is formed for the magnitude of the velocity when both vector components $x$ and $y$ are Gaussian distributed (in the ideal case) with zero mean velocity and equal standard deviation, for example by a random walk process. Thus this kind of a distribution is also a good candidate for the background flow and it might be a good candidate during weak emergences in which the additional flow component still follows more or less a random walk but presumably with higher amplitudes. However, in the case of a strong flux emergence it is highly likely that all velocities in a larger and affected FOV area get directed away from the centre. Therefore, while the velocity amplitudes might still be stronger and weaker, in some way the distribution becomes one dimensional (only radially orientated away from the centre of emergence) and hence the component representing the affected flux emergence pixels follows to a greater extent a Gaussian distribution instead of the Rayleigh one.
A clearer insight could be gained by investigating in the future the formation of several active regions and looking then, with an even higher focus, on the velocity distributions to study these last details, as well as if and how, the secondary component changes.

After the detailed discussion of the results above, we wish to contextualize our work within the larger field of solar physics. The evolution of active regions is an ongoing research field, especially in regards to the build up of magnetic energy for solar eruptions, so-called flares \citep[see e.g.][]{2018MNRAS.477..293K,2018arXiv180100430Y}. This is generally done by having a detailed look into the magnetic field evolution as well as its configuration over time \citep[e.g.][]{2016A&A...596A..69D}. Such investigations are often directly performed by analysis of magnetograms but increasingly commonly also by magnetic field extrapolations \citep[e.g.][]{2008A&A...484..495T}. Another possibility for such analysis comes via simulations and modelling \citep[][]{2012ApJ...757..147C}. It is very clear that for a successful modelling of the process of energy build up, detailed knowledge about the velocity fields transporting the magnetic field above the solar surface, as well as shredding and twisting the field lines, is of great importance. The detailed measurement of flow fields, and derivation of the velocity distributions, are not only important for the evolution of the active regions themselves, but indeed also necessary for large-scale flux transport models such as the advective flux transport (AFT) model \citep[see e.g.][]{2015ApJ...815...90U}. Thus a better knowledge of the velocity fields will also help in the understanding of the global dynamo acting on the Sun. Such flux transport models describe in a simplified way how the magnetic field emerges (e.g. in active regions), is shredded, and then transported via the velocity fields, including the meridional circulation and differential rotation, to the poles, where the fields finally get submerged. Thus a better parametrization of the velocity fields as done for example in this study will be of importance for such modelling efforts. A final interesting field for which this research might yield a new approach is the field of flux emergence studies. Authors like \citet[][]{2015ARep...59..776G} have pointed out that flux emergence can be detected in image data by algorithms using sophisticated multi-fractal spectral analysis and segmentation. On the other hand, we have shown now that not only the structures within the FOV are changed (classically the granulation pattern gets elongated, which can be used within segmentation algorithms) but that the flow field changes remarkably leading to changed velocity distributions. Thus by investigating the flow field statistics, one can also detect and characterize flux emergence events.

\section{Conclusion}
In this paper we looked into the details of the evolving flow patterns in velocity maps during the formation of active region 11190. The used data were obtained from the SDO/HMI instrument as continuum maps and magnetograms to investigate both the continuum proper motions as well as the magnetic field element motions during two emergence events of positive flux leading in consequence to the formation of the active region. Generally we found a high congruence between the plasma flows and the motions of the magnetic elements. This congruence is weakened and distorted during the emergence of new magnetic flux. Moreover, the speeds in the FOV can be fitted in general very accurately with a Rayleigh distribution. Nevertheless, during the flux emergence events the Rayleigh distributions get distorted and at least a secondary flow field component should be added. It is plausible that this component can be either a secondary Rayleigh distribution with a larger width (higher velocities) during the emergence or a Gaussian component. The stronger the emergence, the more likely it is that the secondary component follows a Gaussian distribution, which can be related to the idea that strong emergences lead to radial outflows and, in that sense, to a one-dimensional flow distribution (only a $v_r$ component exists, while normally the flow velocities are made up of a $v_x$ and a $v_y$ component). In order to support the statement about the necessity of a two-component distribution, where the second component is formed due to the strong changes in the flow pattern occurring during the formation of AR 11190, we analysed 
the evolution of a quiet Sun region during the same day. We found that for a quiet Sun flow-field distribution it is sufficient to use a single Rayleigh distribution to fit the speeds distribution (see Fig. \ref{fig12}). It is also possible to observe the temporal evolution of the fitting parameters over 4 hours (see Fig. \ref{fig13}), and conclude that they do not show strong enhancements compared to their general behaviour.


 \begin{acknowledgements}
This research received support by the Austrian Science Fund (FWF) P27800. Jose Iv\'{a}n Campos Rozo and Santiago Vargas Dom\'{\i}nguez acknowledge funding from Universidad Nacional de Colombia research project code 36127: \emph{Magnetic field in the solar atmosphere}. Additional funding was possible through an Odysseus grant of the Fund for Scientific Research-Flanders (FWO Vlaanderen), the IAP P7/08 CHARM (Belspo), and GOA-2015-014 (KU Leuven). This work has also received funding from the European Research Council (ERC) under the European Union's Horizon 2020 research and innovation programme (grant agreement No 724326). Jose Iv\'{a}n Campos Rozo is grateful to the National University, the Research Direction, and the National Astronomical Observatory of Colombia for providing him with a travel grant under the project for new professors and researchers to spend a part of his thesis time at KU Leuven enabling him to collaborate with Prof. T. Van Doorsselaere for this study. He also wishes to acknowledge the whole KU-Leuven University for the space and the academic support provided during his stay in Leuven. HMI/SDO data are courtesy of NASA/SDO and the AIA, EVE, and HMI science teams, and they were obtained from the Joint Science Operation Center (JSOC). Part of this research has been created by using SunPy libraries, an open-source and free community-developed solar data analysis package written in Python. We wish to express our gratitude to the editor of A\&A, and the  anonymous referee for his or her suggestions and comments about the present work, which improved the study considerably.
 \end{acknowledgements}

\bibliographystyle{aa} 
\bibliography{bibpaper} 

\begin{appendix}
\section{Comparison with quiet Sun} 
\begin{figure*}[]
\centering
\includegraphics[scale=0.7]{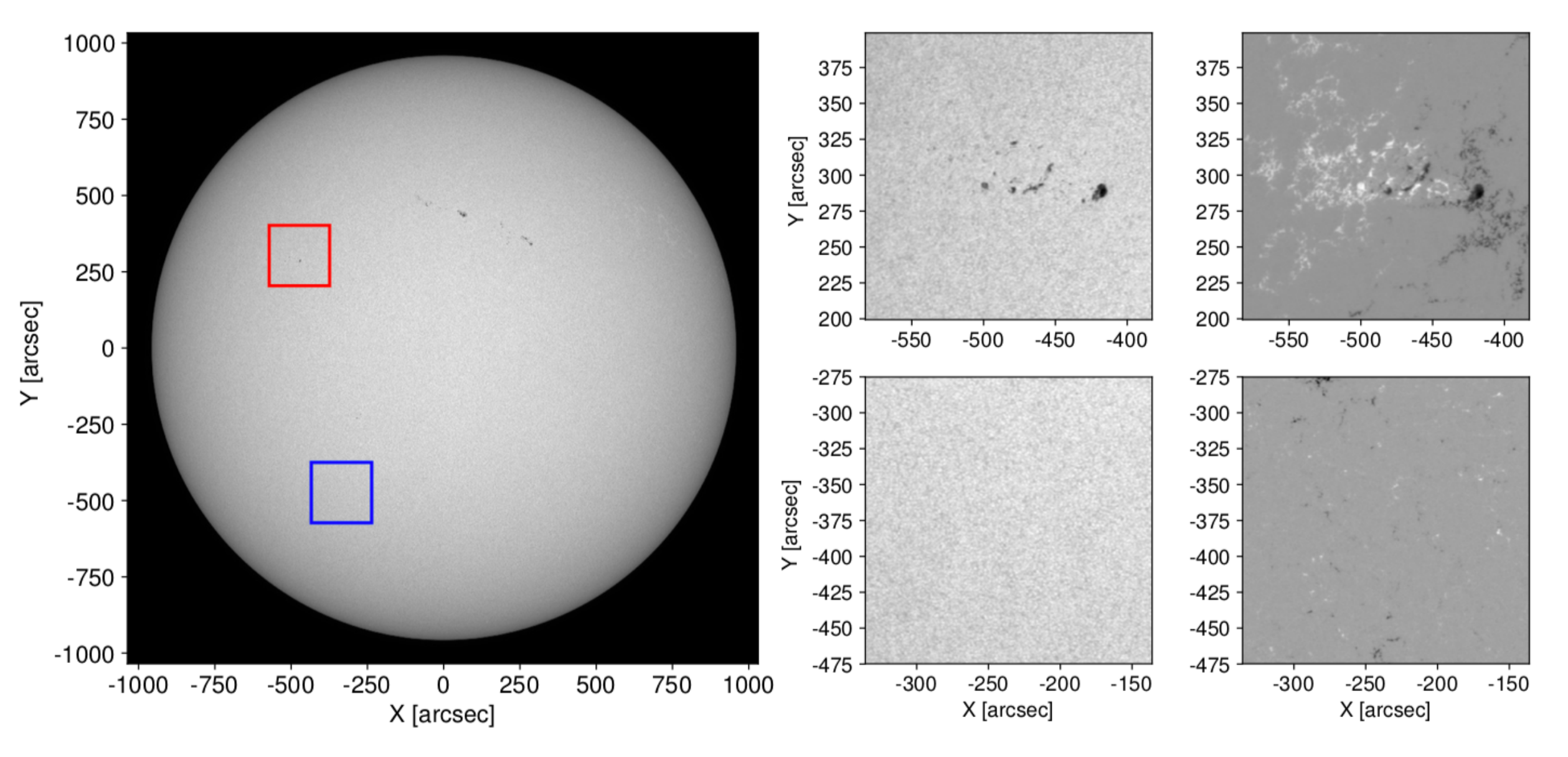}
\caption{Overview of the analysed regions. Left panel: Full disc Sun as seen by SDO/HMI on the day of observations. The two regions of interest are marked by a red square - active region NOAA 11190 - and blue square - quiet Sun region. Right panel, top row from left to right: A continuum map of the NOAA 11190 (red square) followed by the corresponding magnetogram is shown. Bottom row: Same, but for the quiet Sun region. Time shown is April, 11 2011 at 17:45 UT.}
\label{fig11}
\end{figure*}

\begin{figure*}[]
\centering
\includegraphics[scale=0.7]{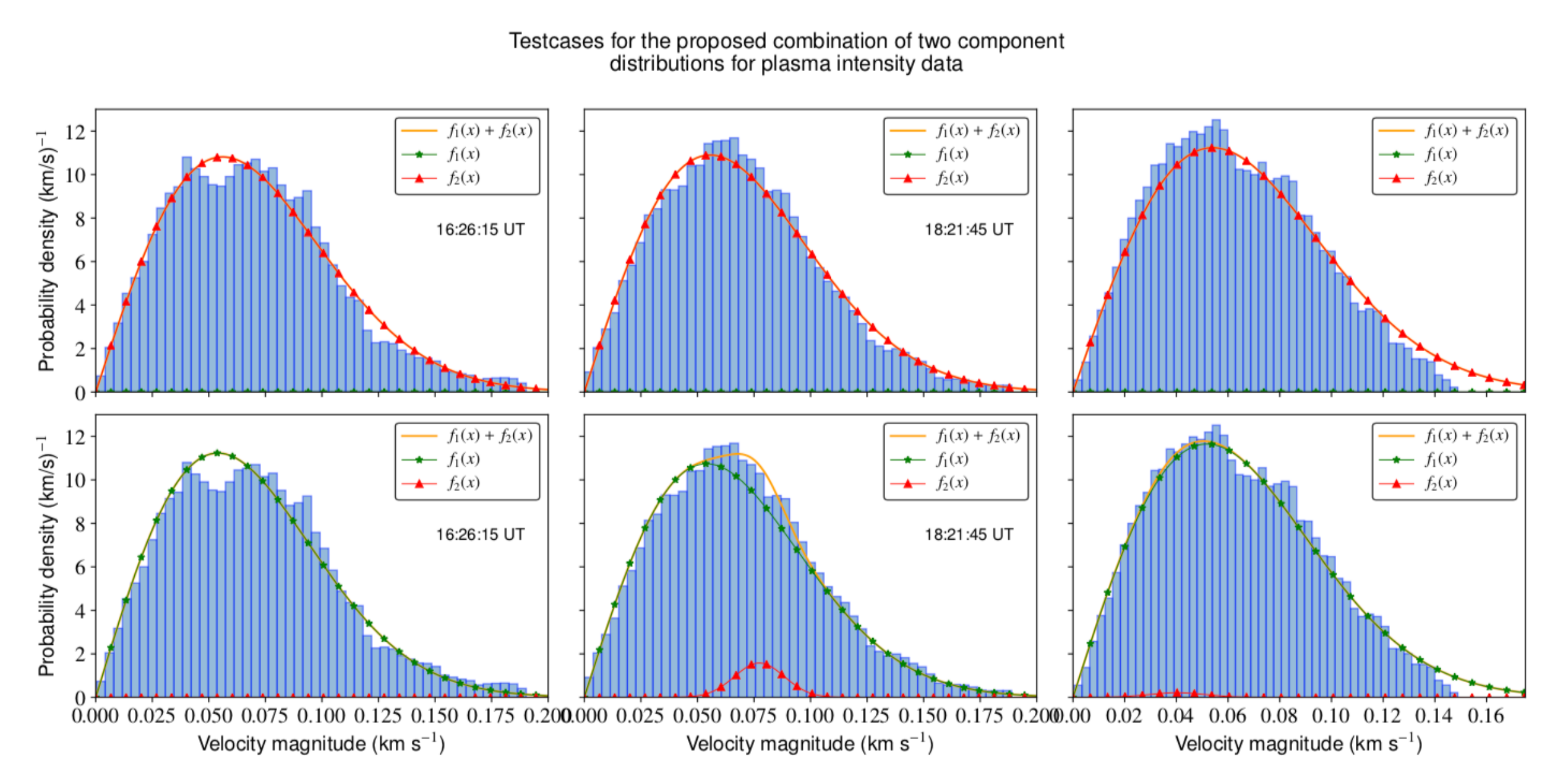}
\caption{Similar to Fig. \ref{fig6}, but for the quiet Sun region as outlined in Fig. \ref{fig11} for three different test cases showing only the distributions obtained from the continuum maps. Thus the arrangement of the panels is as follows: top row shows the histograms of velocity as computed from the continuum maps for three different times applying a double Rayleigh component distribution fit, while the bottom panels give the combination of Rayleigh and Gaussian distribution (all cases obtained from continuum maps).}
\label{fig12}
\end{figure*}

To show the necessity of the combined speed distribution in active regions, we wish to replicate the analysis for a quiet Sun region within the period of our data set (from 16:00 UT to 20:00 UT on the same day).\footnote{An analysis of a plage region would complement this study perfectly, however, due to the considerable size of the current study and the necessary analysis, we postpone such an analysis to a future investigation.} Figure \ref{fig11} shows on the left-hand side the full disc Sun on the day of observations, with the two analysed regions of interest marked by rectangles. The continuum maps and magnetograms of the two regions are shown in the right panels. Clearly the magnetic field activity is very high in the active region, while it is practically non-existent in the quiet Sun (as expected). We applied the LCT algorithm on the chosen quiet Sun region. The data comprise 320 images with the same cadence as described before and a total time of 4 hours. The parameters for the LCT algorithm are the same as in the case for the active region. The principal first outcome can be seen in Fig. \ref{fig12}. Here we show the histograms of the speeds at three different times, which are independent and not related to the AR 11190 analysis. It becomes clear that a single Rayleigh distribution fits very well the whole histogram and it is not necessary, compared to the active region data, to fit the histogram with a more complex two-component distribution. 

\begin{figure*}[]
\includegraphics[scale=0.7]{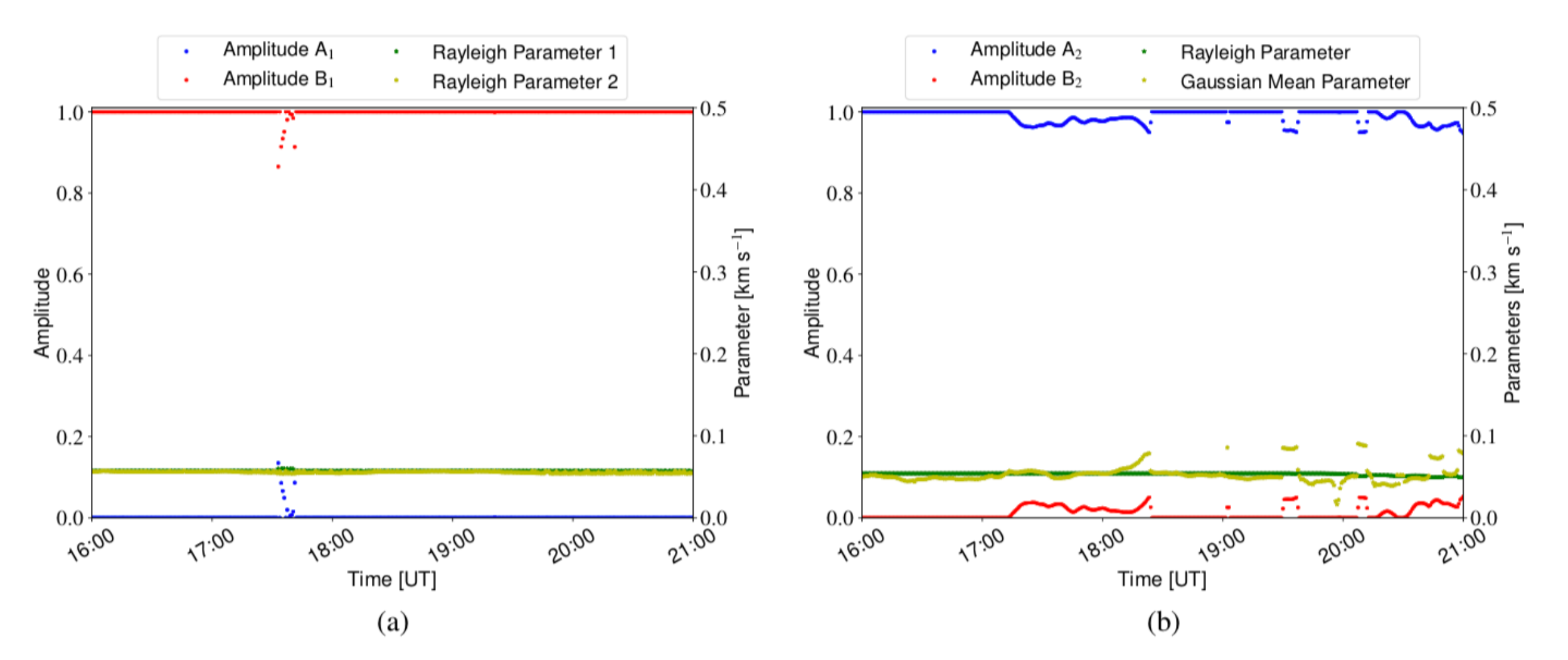}
\caption{Similar to Fig. \ref{fig7}, but for the quiet Sun region as outlined in Fig. \ref{fig11} and only showing the continuum proper motion case (due to quiet Sun conditions, not enough moving magnetic elements are available to obtain velocity distributions).}
\label{fig13}
\end{figure*}

As this could be only a special case for three times, we also replicated Fig. \ref{fig7} for the quiet Sun as shown in Fig. \ref{fig13}. The evolution of the parameters shows no significant events (like strong parameter deviations), except for a few small occasional changes for the combination of Rayleigh distribution with a Gaussian component. Thus, again we can see that within the quiet Sun the expected result was realized, namely, the possibility to create a good single Rayleigh component fit. This is fully understandable as this principal distribution will be formed due to random x/y motions created from the turbulent convection.

\end{appendix}

\end{document}